\documentclass[aps,prl,12pt,amsmath,amssymb]{revtex4-1}

\usepackage{amssymb}
\usepackage{amsmath}
\usepackage{graphicx}
\usepackage{setspace}
\usepackage{xcolor}

\graphicspath {{figures_torque/}}

\newcommand{\Tr}{\operatorname{Tr}}

\renewcommand\Im{\operatorname{Im}}

\DeclareMathOperator{\sign}{sign}

\begin{document}
\title{A Single Gyrotropic Particle as A Heat Engine}

\author{Yu Guo, Shanhui Fan}

\affiliation{
	Department of Electrical Engineering and Ginzton Laboratory, Stanford University, Stanford, California 94305, USA
}


\begin{abstract}
We demonstrate that the system composed of a gyrotropic particle out of thermal equilibrium with vacuum can be regarded as a heat engine. Such a particle, initially at rest, will experience a fluctuation-induced torque and start to rotate, producing mechanical work out from the temperature difference of the particle and its environment. We rigorously prove that the efficiency of the heat engine is tightly bound by the Carnot efficiency. We also predict that, such an engine can be constructed using a heavily-doped semiconductor nanoparticle under magnetic field, and moreover the particle can reach at a steady-state rotating frequency in the order of terahertz solely due to thermal fluctuations.

\end{abstract}

\maketitle

The study of Casimir forces\cite{casimir_attraction_1948,casimir_influence_1948,lifshitz_theory_1956}, including both equilibrium and non-equilibrium forces, has been of fundamental importance since such studies illustrate some of the surprising consequences that quantum and thermal fluctuations can generate in nanoscale systems\cite{lamoreaux_demonstration_1997,mohideen_precision_1998,chen_new_2003,antezza_casimir-lifshitz_2006,obrecht_measurement_2007,capasso_casimir_2007,buhmann_thermal_2008,maia_neto_casimir_2008,rosa_casimir-lifshitz_2008,antezza_casimir-lifshitz_2008,klimchitskaya_casimir_2009,bimonte_scattering_2009,rahi_scattering_2009,munday_measured_2009,rodriguez_achieving_2010,messina_scattering-matrix_2011,sushkov_observation_2011,rodriguez_casimir_2011,kruger_nonequilibrium_2011,kruger_trace_2012,zhao_rotational_2012,reid_fluctuation-induced_2013,bimonte_observing_2015,somers_measurement_2018,jiang_axial_2019,sanders_nanoscale_2019}. While most of the existing works have focused on measuring and computing these forces in a wide variety of material systems and geometries, there are emerging interests in seeking to control these forces such as through dynamic modulation\cite{ma_enhanced_2019}, or to design these forces to achieve specific functionalities such as to achieve stable mechanical equilibrium\cite{kruger_non-equilibrium_2011,chen_nonequilibrium_2016,zhao_stable_2019}.

As a step forward in seeking to explore some of consequences of Casimir forces, in this work we show that non-equilibrium Casimir forces can be harnessed to create a nanoscale heat engine. We consider a spherical nanoparticle made of gyrotropic materials surrounded by vacuum. We show that such a nanoparticle can operate as a heat engine. In the non-equilibrium scenario when its temperature is different from that of the surrounding environment, due to thermal fluctuations the particle will start rotating and eventually reach a steady-state angular frequency. Thus it converts thermal energy from random current fluctuation to kinetic energy in the rotation of the entire body. We further show that the efficiency of such conversion is tightly bounded by the Carnot limit. Our work indicates that non-equilibrium Casimir forces may be exploited for energy harvesting at nanoscale.

Our work is related to a few recent developments. It has been pointed out that the thermal radiation of a non-reciprocal object, such as a gyrotropic particle\cite{ott_circular_2018} or slab\cite{khandekar_thermal_2019}, or a topological insulator thin film\cite{maghrebi_fluctuation-induced_2019}, carries angular momentum. And hence there is a back-action torque on the object when it is out of thermal equilibrium with the environment. The studies of Ref.~\cite{ott_circular_2018,khandekar_thermal_2019,khandekar_circularly_2019,maghrebi_fluctuation-induced_2019} all assume that these objects are stationary. In Ref.~\cite{manjavacas_vacuum_2010,manjavacas_thermal_2010}, it was noted that a particle, when rotating at a constant angular velocity, can experience vacuum friction and hence transform part of the mechanical energy to thermal energy. Building upon these works we provide a unified treatments of both the effects mentioned above, including the effects of the back-action torque on the motion of the particle. The notion that non-equilibrium Casimir forces can be used to construct a heat engine has not been previously recognized.

\begin{figure}
	\centering
	\includegraphics[width=0.3\textwidth]{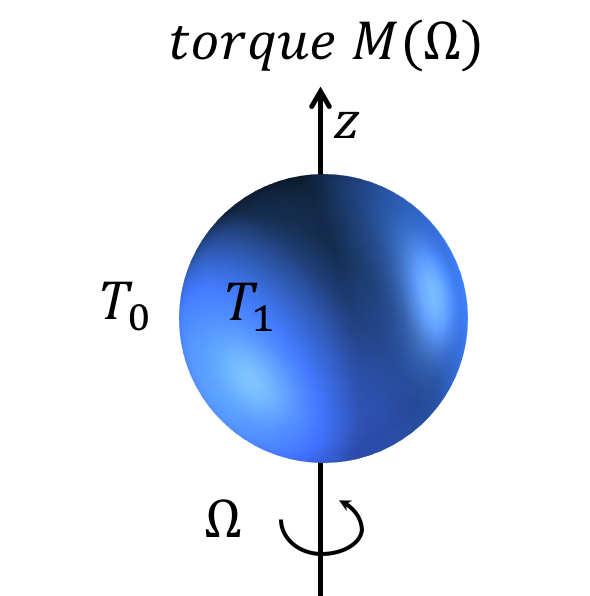}
	\caption{Schematic of the structure considered in this paper. The structure consists of a gyrotropic nano-sphere at a temperature $T_1$, embedded in vacuum at a temperature $T_0$. We show that this structure can operate as a heat engine. \label{fig:schematic}}
\end{figure}

We consider a single particle in vacuum. The temperature of the particle is $T_1$ and that of the vacuum is $T_0$. The permittivity tensor of the particle is given by 
\begin{align}
\epsilon=\begin{pmatrix}
\epsilon_{xx} & \epsilon_{xy} &  0 \\
\epsilon_{yx} & \epsilon_{yy} & 0 \\
0 & 0 &\epsilon_{zz}
\end{pmatrix}
\end{align}
We assume that the size of the particle is sufficiently small compared to the wavelength of thermal photons at the relevant temperatures. Hence the particle can be described by its frequency dependent polarizability $\alpha(\omega)$:
\begin{align}
\alpha=\begin{pmatrix}
\alpha_{xx} & \alpha_{xy} &  0 \\
\alpha_{yx} & \alpha_{yy} & 0 \\
0 & 0 &\alpha_{zz}
\end{pmatrix}
\end{align}
Denote the upper $2\times 2$ principal minor of the matrix $\alpha$ by $\alpha_\perp$, that is, $\alpha_\perp=[\alpha_{xx}, \alpha_{xy}; \alpha_{yx}, \alpha_{yy}]$. We expand $\alpha_\perp$ as
\begin{align}\label{eq:alpha_perp}
\alpha_\perp=\alpha_0I_2+\alpha_1\sigma_x+\alpha_2\sigma_y+\alpha_3\sigma_z,
\end{align}
where $I_2$ is the 2$\times$2 identity matrix and $\sigma_x$, $\sigma_y$, $\sigma_z$ are the standard Pauli matrices. From $\alpha_\perp(-\omega)=\alpha_\perp^*(\omega)$, we have
\begin{align} \label{eq:alpha}
\alpha_0(-\omega)=\alpha_0^*(\omega), \: \alpha_2(-\omega)=-\alpha_2^*(\omega)
\end{align}
Since $\sigma_y$ is antisymmetric whereas all the other three matrices in Eq.~(\ref{eq:alpha_perp}) are symmetric, a particle is reciprocal if and only if $\alpha_2=0$. 
In the following, we denote 
\begin{align}\label{eq:g_perp}
g_\perp(\omega)=\Im(\alpha_0(\omega)+\alpha_2(\omega)).
\end{align}

With a rotation frequency $\Omega$ around the $z$-axis, the total torque on the particle is given by\cite{torque_SM_1} 
\begin{align} \label{eq:torque_exact}
\begin{split}
M&=-\frac{4\hbar}{3\pi c^3}\int_{-\infty}^\infty{d\omega}\omega^3 g_\perp(\omega_-) (n_1(\omega_-)-n_0(\omega))
\end{split}
\end{align}
where $\omega_-=\omega-\Omega$, $n_i(\omega)=\frac{1}{e^{\hbar\omega/(k_BT_i)}-1}+\frac{1}{2}=-n_i(-\omega)$. 
The net power radiated by the particle to free space is given by $P=P_\perp+P_\parallel$, where \cite{torque_SM_2}
\begin{align}\label{eq:power_perp}
\begin{split}
P_\perp&=\frac{4\hbar}{3\pi c^3}\int_{-\infty}^\infty{d\omega}\omega^4 g_\perp(\omega_-) (n_1(\omega_-)-n_0(\omega))
\end{split}
\end{align}
\begin{align}\label{eq:power_para}
P_\parallel=\frac{2\hbar}{3\pi c^3}\int_{-\infty}^\infty{d\omega}\omega^4\Im\alpha_{zz}(\omega)(n_1(\omega)-n_0(\omega))
\end{align}
Note that $P_\perp$ depends on the rotation frequency $\Omega$ and can be negative for $T_1>T_0$, 
while $P_\parallel$ is independent of $\Omega$ and is always non-negative for $T_1>T_0$.

The theory above provides the dependency of fluctuation-induced torque $M(\Omega)$ with respect to the rotation frequency $\Omega$. In order for such a nanoparticle to operate as a heat engine, the engine needs to be able to start, thus the particle must experience a non-zero torque when it is stationary, i.e. $M(0)\neq 0$. Examining Eq.~(\ref{eq:torque_exact}), we see that in order for $M(0)\neq 0$, the system must satisfy the following two conditions: (1) It is out of equilibrium. The torque $M(0)=0$ if $T_1=T_0$. This is required by the second law of thermodynamics: one should not be able to construct an engine at equilibrium. (2) The nanoparticle must be made of non-reciprocal materials. At $\Omega=0$, the $\alpha_0 (\omega)$ term in Eq.~(\ref{eq:torque_exact}) cancels after the integration, as can be seen from Eq.~(\ref{eq:alpha}). Thus, $M(0)\neq 0$ implies that $\Im \alpha_2 (\omega)\neq 0$.

In the presence of the torque $M(\Omega)$, the dynamics of the nanoparticle is described by $I\frac{\mathrm{d} \Omega}{\mathrm{d}t}=M(\Omega)$, with $I$ being the moment of inertia of the particle. Suppose at $t=0$ the particle is stationary, since $M(\Omega=0)\neq 0$, the particle will start to rotate. The rotation will accelerate as long as $\Omega M(\Omega)> 0$. On physical ground one would expect that the torque tends to slow down the rotation if the particle rotates sufficiently fast, that is, $\Omega M(\Omega)<0$ when $|\Omega|$ is sufficiently large. Thus there should exist a steady-state rotation frequency $\Omega_0$ where $M(\Omega_0)=0$ and $\Omega_0M(0)>0$. The rotation of the particle keep accelerating until $\Omega=\Omega_0$. In the process for the particle to reach steady-state, we have $\Omega M(\Omega) \geq 0$, that is, the particle is producing mechanical energy out of the temperature difference between the environment and itself. Hence our structure can be seen as a heat engine.



\textit{Efficiency of heat engine}.
One important property of a heat engine is its efficiency. In our system, at a given rotation frequency $\Omega$, the mechanical work generated by the system is $\Omega M$, the heat flow from the hot body to the cold body is $\hat{P}=\sign(T_1-T_0)P$, since in our definition the radiated power $P$ (Eqs.~\ref{eq:power_perp}, \ref{eq:power_para}) is the net power from the particle to environment. $\hat{P}_\perp$, $\hat{P}_\parallel$ are defined likewise and the latter is always non-negative. By definition, the efficiency of our heat engine is given by

\begin{align}\label{eq:efficiency}
\eta=
\begin{cases}
{\Omega M}/{(\Omega M+\hat{P})} & T_1>T_0\\
{\Omega M}/{\hat{P}} & T_1<T_0\\
\end{cases}
\end{align}
We also define $\tilde{\eta}$ by replacing $\hat{P}$ with $\hat{P}_\perp$ in Eq.~\ref{eq:efficiency}, since $\hat{P}_\parallel$ is independent of rotation and hence does not contribute to producing mechanical work. In addition, as $\hat{P}_\parallel$ is always non-negative, we have $
\eta\leq \tilde{\eta}$ and the equality holds when $\hat{P}_\parallel=0$. The definitions of efficiency $\eta$, $\tilde{\eta}$ are meaningful only if $\Omega M \geq 0$.

We first show that with a proper choice of the polarizability tensor, the efficiency ${\eta}$ of our heat engine can reach the Carnot efficiency limit $\eta_C=\frac{|T_1-T_0|}{\max(T_1,T_0)}$. For this purpose, we assume that
\begin{align}
&\Im \alpha_{zz}(\omega) = 0 \label{eq:condition_alpha_zz}\\
g_\perp &(\omega)=g\delta(\omega-\omega_0) \label{eq:condition_g}
\end{align}
where $\omega_0\neq 0$. From Eq.~(\ref{eq:g_perp}), this ideal $g_\perp(\omega)$ can be constructed with $\Im(\alpha_0) = g(\delta(\omega-\omega_0) - \delta(\omega+\omega_0))/2$, $\Im(\alpha_2) = g(\delta(\omega-\omega_0) + \delta(\omega+\omega_0))/2$. Both $\alpha_0$ and $\alpha_2$ here satisfy the constraint of Eq.~(\ref{eq:alpha}).

Under these conditions Eqs.~(\ref{eq:condition_alpha_zz}) and (\ref{eq:condition_g}), we have $P_\parallel=0$ from Eq.~\ref{eq:power_para}. Also, from Eq.~\ref{eq:torque_exact}, the steady-state frequency $\Omega_0$ where $M(\Omega_0)=0$ should satisfy $n_1(\omega_0)=n_0(\omega_0+\Omega_0)$. Hence
\begin{align}\label{eq:steady_Omega0}
\Omega_0=\omega_0(\frac{T_0}{T_1}-1).
\end{align}
Here we note that $\Omega_0$ changes sign between $T_1 < T_0$ and $T_1 > T_0$, indicating the change of rotation direction between these two scenarios. For a rotation frequency $\Omega$, having the same sign as $\Omega_0$, and with $0 < |\Omega|<|\Omega_0|$, the efficiency ${\eta}$ is given by,
\begin{align}\label{eq:efficiency_Delta}
{\eta}=
\begin{cases}
-\frac{\Omega}{\omega_0}  & T_1>T_0\\
\frac{\Omega}{(\omega_0+\Omega)} & T_1<T_0\\
\end{cases}.
\end{align}
In either case, when the magnitude of rotation frequency $|\Omega|$ is changing from 0 to $|\Omega_0|$, the efficiency ${\eta}$ increases from 0 to the Carnot efficiency $\eta_C$. In addition, when $\Omega$ is approaching $\Omega_0$ and hence the efficiency $\eta$ is approaching the Carnot limit $\eta_C$, the torque and the power are both approaching zero, as expected for a heat engine operating near the Carnot limit.

We next rigorously prove that, for any physical polarizability tensor $\alpha(\omega)$, when the system is generating mechanical energy, i.e., $\Omega M >0$, the radiative power $\hat P$ is positive, and the efficiency $\eta$ is bound by the Carnot efficiency limit, 
\begin{align}\label{eq:efficiency_limit}
\eta\leq \tilde{\eta}\leq \eta_C = \frac{|T_1-T_0|}{\max(T_1,T_0)}.
\end{align}
To do this, we first prove that 
\begin{align}\label{eq:efficiency_proof}
T_0 \Omega M \leq |T_1-T_0|\hat{P}_\perp = (T_1-T_0)P_\perp
\end{align}
holds true for any rotation frequency $\Omega$ and temperatures.
By using Eqs.~(\ref{eq:torque_exact}) and (\ref{eq:power_perp}), and changing the integration variable from $\omega$ to $\omega+\Omega$, we have 
\begin{align}\label{eq:efficiency_int}
\begin{split}
&(T_1-T_0)P_\perp-T_0\Omega M=\frac{4\hbar}{3\pi c^3}\int_{-\infty}^\infty{d\omega}(\omega+\Omega)^3 \\ 
&\qquad(T_1(\omega+\Omega)-T_0\omega)g_\perp(\omega)(n_1(\omega)-n_0(\omega+\Omega))
\end{split}
\end{align}
First, we note that $\omega\frac{\alpha(\omega)-\alpha^\dagger(\omega)}{2i}$ must be a semi-positive definite matrix, because $\alpha(\omega)$ is a linear response function of the static particle which is a passive system. As a result, $\omega\Im(\alpha_0(\omega))\geq |\omega\Im(\alpha_2(\omega))|\geq 0$ \cite{torque_SM_3}. It follows that $\omega g_\perp(\omega)=\omega\Im(\alpha_0(\omega)+\alpha_2(\omega))\geq 0$.
Second, one can show mathematically that $\omega(\omega+\Omega)(T_1(\omega+\Omega)-T_0\omega)(n_1(\omega)-n_0(\omega+\Omega))\geq 0$ for any $\omega$ and $\Omega$. Therefore, the integrand in Eq.~(\ref{eq:efficiency_int}) is non-negative at every frequency, which completes the proof of Eq.~(\ref{eq:efficiency_proof}). As a result, if $\Omega M>0$, $\hat{P}_\perp>0$ and from Eq.~(\ref{eq:efficiency_proof}) one can show mathematically that $\tilde{\eta}\leq \eta_C$. Since $\hat{P}_\parallel\geq 0$, one finds $\eta\leq \tilde{\eta}$ when $\Omega M>0$, which completes our proof of Eq.~(\ref{eq:efficiency_limit}). This proof of the efficiency bound is independent of the explicit formalism of polarizability, which can include the radiation correction and magnetic polarizability.

The same system as discussed above can also operate as a heat pump, when one applies an external torque onto the nanoparticle, such that the nanoparticle rotate at a frequency above $\Omega_0$. With $\Omega>\Omega_0$, one can show that the net heat flow goes from the cold side to the hot side, and the efficiency of the heat pump approaches the Carnot limit at $\Omega\to \Omega_0$, under the same conditions as described by Eqs.~(\ref{eq:condition_alpha_zz}) and (\ref{eq:condition_g}).

It what follows we examine such heat engine in more details, by assuming that the sphere is made of heavily doped semiconductor. A heavily doped semiconductor, in the presence of an external magnetic field along the $z$-direction, has a permittivity described as\cite{zhu_near-complete_2014}:
\begin{align}\label{eq:eps_model}
\begin{split}
\epsilon= 
\begin{pmatrix}
\epsilon_\infty-\frac{\omega_p^2(\omega+i\gamma)}{\omega[(\omega+i\gamma)^2-\omega_c^2]} & \frac{i\omega_p^2\omega_c}{\omega[(\omega+i\gamma)^2-\omega_c^2]} & \\
-\frac{i\omega_p^2\omega_c}{\omega[(\omega+i\gamma)^2-\omega_c^2]} & \epsilon_\infty-\frac{\omega_p^2(\omega+i\gamma)}{\omega[(\omega+i\gamma)^2-\omega_c^2]} & \\
& & \epsilon_{zz}
\end{pmatrix}
\end{split}
\end{align}
where $\epsilon_{zz}=\epsilon_\infty-\frac{\omega_p^2}{\omega(\omega+i\gamma)}$. Below we refer to this model as the \textit{magnetized plasma model}. With this model, since $\Im \epsilon_{zz}(\omega) \neq 0$, the nanoparticle has a polarizability with $\Im \alpha_{zz}(\omega) \neq 0$, thus such a particle cannot operate at Carnot limit. In order to bridge the Carnot efficiency limit and the performance of more realistic materials, below we first consider a model that has the same form of permittivity as Eq.~(\ref{eq:eps_model}) except for a frequency-independent and real $\epsilon_{zz}$. Below for concreteness we set $\epsilon_{zz}=2$, and $\epsilon_\infty=1$.
One can create such a medium, starting with a material with free-electron response only in the in-plane directions, and then apply a magnetic field in the $z$-direction. Since in the absence of external magnetic field such a material behaves as a hyperbolic material\cite{cortes_quantum_2012,poddubny_hyperbolic_2013,jahani_all-dielectric_2016}, below for brevity we refer to this model as the \textit{magnetized hyperbolic model}. The use of this model allows us on one hand to study how the efficiency approaches the Carnot limit as one varies the parameter of the model, and on the other hand makes connection to the more realistic magnetized plasma model that may be demonstrated experimentally.

We next show that the magnetized hyperbolic model, with the right choice of parameters, can lead to an engine operating at the Carnot limit. For a spherical particle, its electric polarizability is given by 
\begin{align}
\alpha=\frac{V}{4\pi}(\frac{1}{3}I_3+(\epsilon-I_3)^{-1})^{-1}
\end{align}
where $V$ is the volume of the sphere.
For the magnetized hyperbolic model, we find that 
\begin{align*}
g_\perp(\omega)=\frac{V}{4\pi}\frac{9\gamma\omega\omega_p^2}{(\omega_p^2-3\omega^2+3\omega\omega_c)^2+(3\omega\gamma)^2}
\end{align*}
In the limit of $\gamma \to 0^+$ \cite{torque_SM_4},
\begin{align}
g_\perp(\omega)=\frac{V\omega_p^2}{4\sqrt{\omega_c^2+4\omega_p^2/3}}(\delta(\omega-\omega_{1})-\delta(\omega+\omega_{2}))
\end{align}
where $\omega_1=\frac{\sqrt{\omega_c^2+4\omega_p^2/3}+\omega_c}{2}$ and $\omega_2=\frac{\sqrt{\omega_c^2+4\omega_p^2/3}-\omega_c}{2}$. And therefore, 
\begin{align}\label{eq:torque_zero_loss}
\begin{split}
M &= -s\big[(\omega_{1}+\Omega)^3(n_1(\omega_{1})-n_0(\omega_{1}+\Omega)) \\
&\quad - (\omega_{2}-\Omega)^3(n_1(\omega_{2})-n_0(\omega_{2}-\Omega))\big] 
\end{split}
\end{align}
\begin{align}\label{eq:power_zero_loss}
\begin{split}
P_\perp &= s\big[(\omega_{1}+\Omega)^4(n_1(\omega_{1})-n_0(\omega_{1}+\Omega)) \\
&\quad +(\omega_{2}-\Omega)^4(n_1(\omega_{2})-n_0(\omega_{2}-\Omega))\big]
\end{split}
\end{align}
where $s=\frac{V\hbar\omega_p^2}{3\pi c^3\sqrt{\omega_c^2+4\omega_p^2/3}}$. With a large $\omega_c$, the contributions from the two resonances at $\omega_1$ and $\omega_2$ can be very different, and hence the system can approach the condition of Eqs.~(\ref{eq:condition_alpha_zz}-\ref{eq:condition_g}), where a single resonance peak dominates the response as is required for reaching the Carnot limit.

\begin{figure}
\centering
\includegraphics[width=0.75\textwidth]{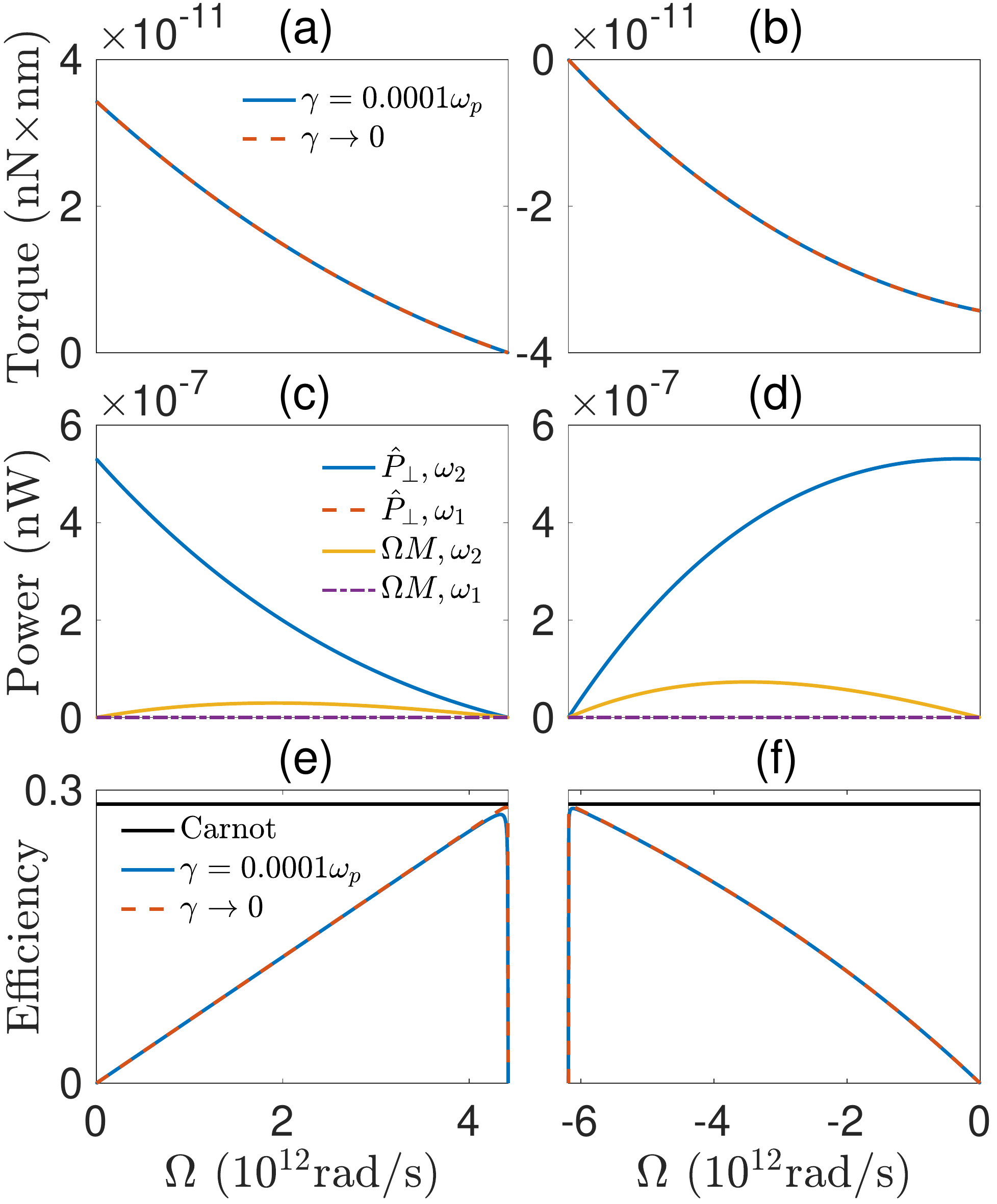}
\caption{Various properties of the structure shown in Fig.~\ref{fig:schematic} as a function of rotating frequency. Here we assume $\omega_c=2\omega_p$, $\gamma=0.0001\omega_p$, $a=100$nm. (a, c, e) $T_1$=70K, $T_2$=50K. (b, d, f) $T_1$=50K, $T_2$=70K. \label{fig:carnot}}
\end{figure}

\begin{figure}
\centering
\includegraphics[width=0.75\textwidth]{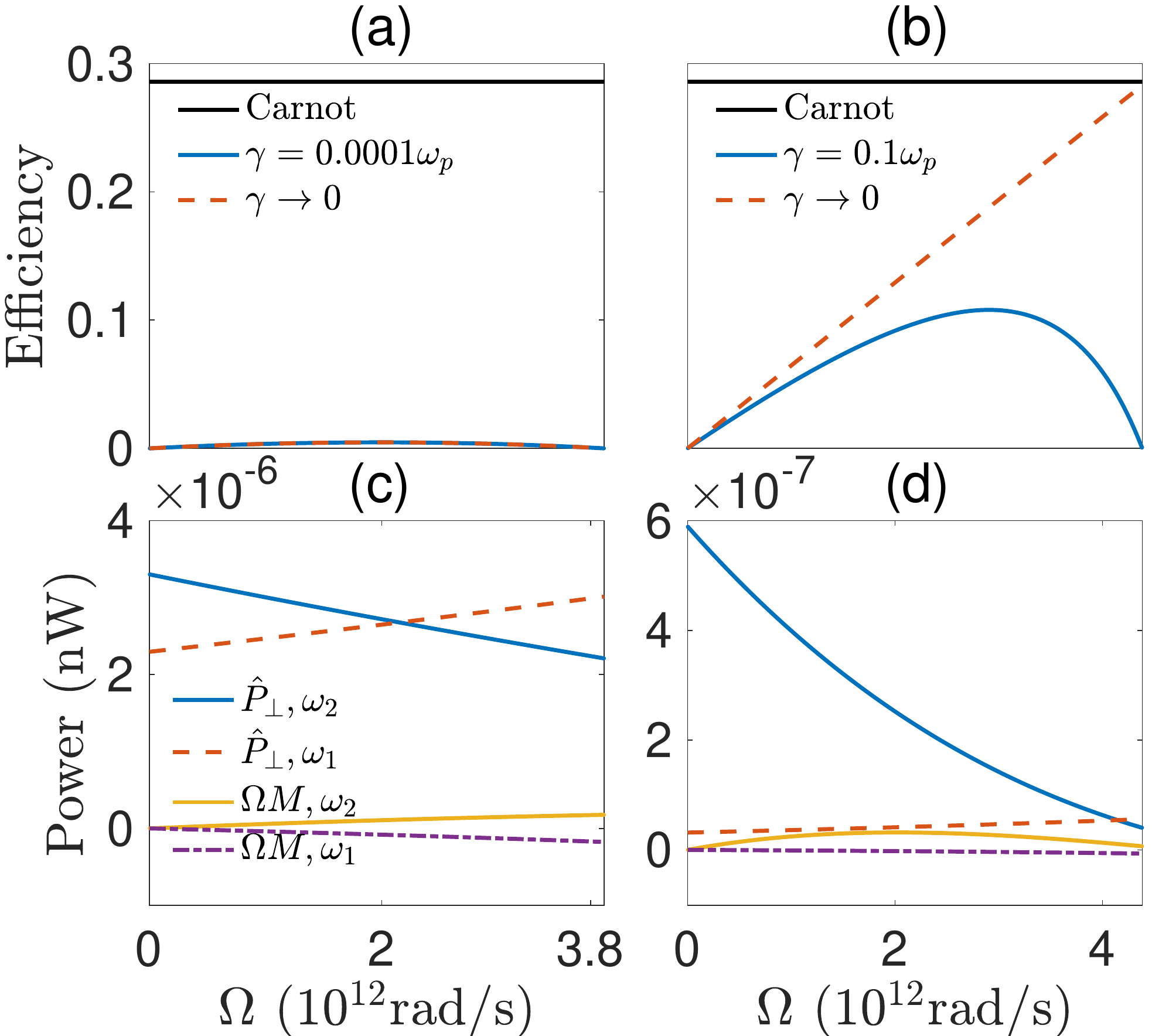}
\caption{Various properties of the structure shown in Fig.~\ref{fig:schematic} as a function of rotating frequency. Here we assume $T_1$=70K, $T_2$=50K, $a=100$nm. (a,c) $\omega_c=0.1\omega_p$, $\gamma=0.0001\omega_p$. (b,d) $\omega_c=2\omega_p$, $\gamma=0.1\omega_p$. \label{fig:nonideal}}
\end{figure}

\begin{figure}
\centering
\includegraphics[width=0.75\textwidth]{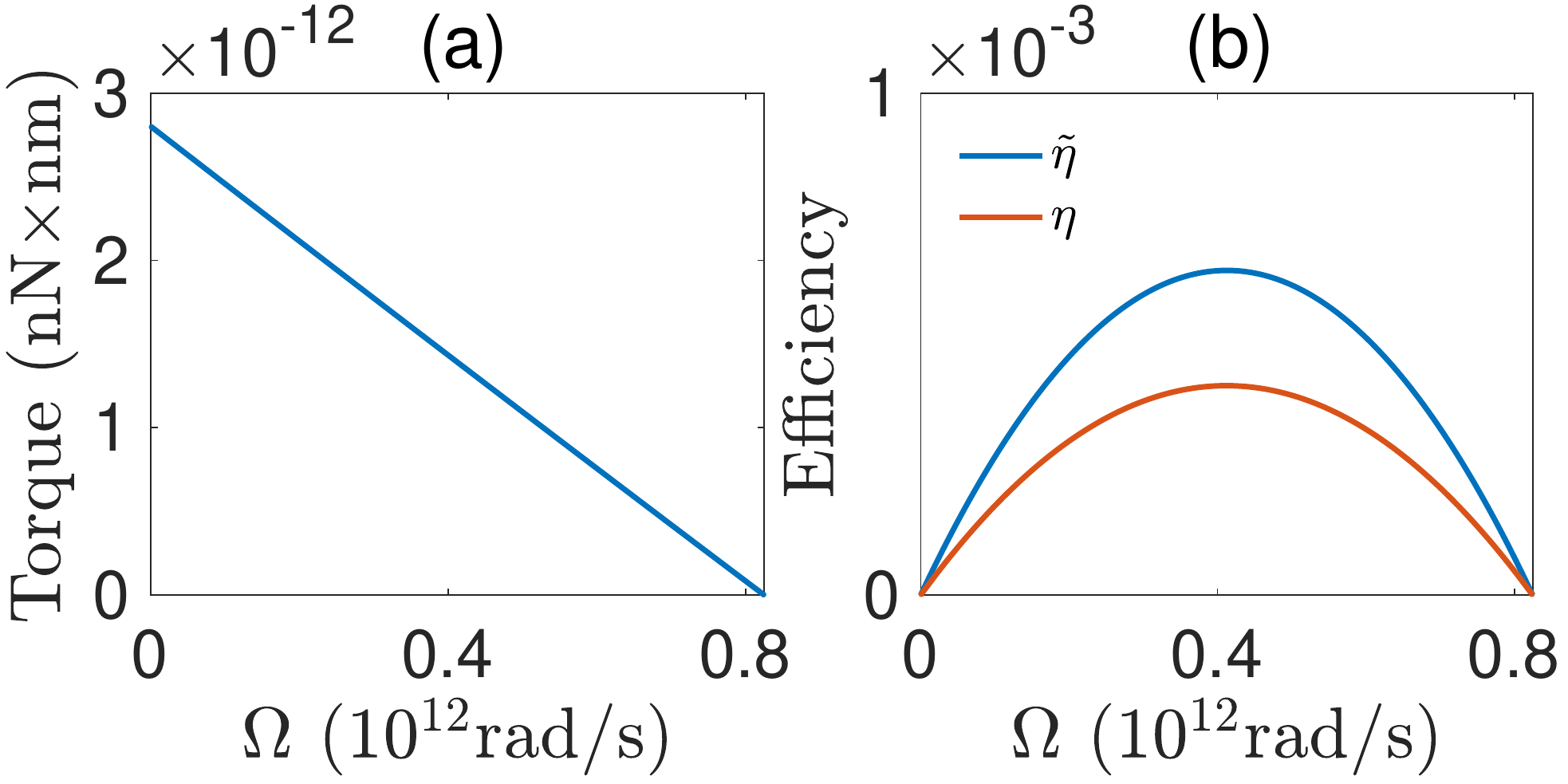}
\caption{(a) Torque versus rotation frequency. (b) Efficiency versus rotation frequency. Here we assume a InSb nanosphere under static external magnetic field. The parameters are $\epsilon_\infty=12.37$, $\omega_p=1.335\times 10^{14}$rad/s, $\omega_c=1.407\times 10^{13}$rad/s, $\gamma=1.922\times 10^{12}$rad/s, $a=100$nm, $T_1$=70K and $T_2$=50K.\label{fig:InSb}}
\end{figure}

For numerical simulations, we choose $\omega_p=10^{14}$rad/s, and the radius of the sphere $a=100$nm. In Fig.~\ref{fig:carnot}, we demonstrate that we can approach the Carnot efficiency with the magnetized hyperbolic model. We choose $\omega_c=2\omega_p$ and $\gamma=0.0001\omega_p$. In the left panel, the temperatures are $T_1$=70K and $T_0$=50K; in the right panel, the temperatures are flipped, that is, $T_1$=50K and $T_0$=70K. In Fig.~\ref{fig:carnot}(a, b), we plot the torque as a function of the rotation frequency $\Omega$. At $\Omega=0$, the torque is positive in Fig.~\ref{fig:carnot}(a) while negative in Fig.~\ref{fig:carnot}(b), so that a particle that is initially stationary will start rotating in opposite directions. In either case, the torque decreases as a function of the magnitude of the rotating frequency. The frequency where the torque reaches zero defines the steady-state rotating frequency $\Omega_0$ of the particle. For both cases of $T_1 > T_2$ and $T_1 < T_2$ there is a range of $\Omega$ for which $M\Omega > 0$, and hence work is generated from the temperature gradient. In both Fig.~\ref{fig:carnot}(a) and (b), the
results of exact computation with Eqs.~(\ref{eq:torque_exact}) and (\ref{eq:power_perp}) agree well with low loss limit of Eqs.~(\ref{eq:torque_zero_loss}) and (\ref{eq:power_zero_loss}).

In Fig.~\ref{fig:carnot}(c) and (d), we plot the mechanical and thermal power as a function of the rotating frequency. Here, we separately plot the contributions from the two resonances at $\omega_1$ and $\omega_2$. We see that the contribution from $\omega_2$ dominates while the contribution from $\omega_1$ is negligible. At a large $\omega_c$ there is a large difference between $\omega_1$ and $\omega_2$. With our choice of temperatures only one of these resonances have significant thermal excitation. Consequently, the system behavior approaches that as described by Eqs.~(\ref{eq:condition_alpha_zz}) and (\ref{eq:condition_g}). In Fig.~\ref{fig:carnot}(e) and (f), we plot the numerically calculated efficiency. Comparing the efficiency to the formula of Eq.~(\ref{eq:efficiency_Delta}) where we set $\omega_0=\omega_2$, for both choices of temperatures we see excellent agreement between the theory and the numerical results. The efficiency indeed gets very close to the Carnot limit as $\Omega$ approaches the steady-state frequencies at either cases. The power generated (Fig.~\ref{fig:carnot}(c) and (d)) goes towards zero as the efficiency approaches Carnot limit. The efficiency $\eta$ is zero at the steady-state frequency $\Omega_0$ because at $\Omega_0$ the torque $T$ is zero while the power $\hat{P}_\perp$ is positive due to the small but non-zero contribution from the $\omega_1$ term, which is in contrary to Eq.~(\ref{eq:efficiency_Delta}) where both the torque and the power are zero at the steady-state frequency.

The results of Fig.~\ref{fig:carnot} demonstrate that a heat engine approaching Carnot efficiency can be achieved using a sphere as described by the magnetized hyperbolic model in the limits of large $\omega_c$ and small $\gamma$. In Fig.~\ref{fig:nonideal}, we consider the influence on the performance of such heat engine as we deviate from such limits. Fig.~\ref{fig:nonideal}(a) shows the efficiency in the case where $\omega_c$ is reduced to 0.1$\omega_p$ whereas $
\gamma$ is maintained at $0.0001\omega_p$. We see that the efficiency is significantly reduced compared with the case shown in Fig.~\ref{fig:carnot}c, where $\omega_c=2\omega_p$. In Fig.~\ref{fig:nonideal}(c) we show the radiated power and the mechanical power, separating the contributions from the two resonances $\omega_1$ and $\omega_2$. In contrast to the case shown in Fig.~\ref{fig:carnot}e, here there are significant contributions from both resonances. Their contributions to the mechanical power, however, have opposite signs. And hence the total generated mechanical power and the efficiency is far lower as compared with the case in Fig.~\ref{fig:carnot}.

Fig.~\ref{fig:nonideal}(b) shows the case where $\omega_c$ is maintained at 2$\omega_p$, while we increase $\gamma$ to 0.1$\omega_p$. The efficiency degrades from the ideal case, since the thermal emission is now over a broader range of wavelengths. But the decrease in efficiency is not nearly as drastic as compared to the case in Fig.~\ref{fig:nonideal}(a). In Fig.~\ref{fig:nonideal}(d), we plot the radiated power and the mechanical power in this case, again separating the contributions from the two resonances $\omega_1$ and $\omega_2$. Here, we integrate Eqs.~(\ref{eq:torque_exact}) and (\ref{eq:power_perp}) around the two resonances $\omega_1$ and $\omega_2$ to approximate their individual contributions. We see that the contributions from $\omega_2$ remains dominant. To reach an efficiency that is comparable to Carnot efficiency, a large $\omega_c$ is therefore essential. In the magnetized plasma model, $\omega_c$ is proportional to the strength of the external magnetic field, and measures the strength of non-reciprocity. Therefore, our results here indicate that reciprocity-breaking is essential for the operating of such heat engine.

In connection with the behavior of such heat engine when the particle is described by the magnetized hyperbolic model,  we now consider the case where the particle is described by the magnetized plasma model. In this case, we assume parameters $\epsilon_\infty=12.37$, $\omega_p=1.335\times 10^{14}$rad/s, $\omega_c=1.407\times 10^{13}$rad/s, $\gamma=1.922\times 10^{12}$rad/s, $a=100$nm. These parameters corresponds to n-doped InSb, with a dopant level of $1.4\times 10^{17}$/cm$^3$, and under a magnetic field of 2T\cite{zhu_near-complete_2014}. We consider the case where $T_1$=70K and $T_2$=50K. Fig.~\ref{fig:InSb}(a) shows the torque as a function of rotating frequency. We see similar behavior as the idealized case of Fig.~\ref{fig:carnot}(a). And hence the system behaves as a heat engine. Compared with the idealized case shown in Fig.~\ref{fig:carnot}, the torque and the efficiency decreases by 1 and 2 orders of magnitude, respectively. This arises primarily due to the smaller strength of non-reciprocity, as well as a broader linewidth of the resonance, in the InSb system, as we have discussed above in Fig.~\ref{fig:nonideal}. In addition, for this system $P_\perp$ is no longer zero, and hence there are heat flows from the particle to the environment which does not contribute to the torque, which further decreases the efficiency. The impact of $P_\perp$ is shown in Fig.~\ref{fig:InSb}(b), where we compare the efficiency of the system $\eta$, with $\hat{\eta}$ for which the $P_\perp$ is excluded. We also note that in this case the efficiency peaks at a frequency substantially lower than the steady-state frequency $\Omega_0$, since as $\Omega$ goes to $\Omega_0$ the mechanical power approaches zero but the radiative power does not. 

\textit{Concluding remarks}.
We demonstrate theoretically and numerically that a nanoparticle made of gyrotropic media in vacuum can operate as a heat engine due to non-equilibrium Casimir forces. We prove that the efficiency of such a heat engine is bound by the Carnot limit. We also highlight the important considerations on material dielectric response in order to approach the Carnot limit. In practice such an engine can be constructed using an InSb nanoparticle under magnetic field, even though the efficiency for the InSb nanoparticle is much smaller compared to the Carnot limit. Our theory also predicts that the InSb particle, initially at rest, can reach a steady-state rotating frequency in the order of $10^{12}$rad/s due to the torque from thermal fluctuations.

For a static nanoparticle with a size in the deep sub-wavelength regime, non-reciprocity in its material dielectric response is necessary for the existence of the torque. Non-reciprocity can be achieved with the use of external magnetic field, as we considered in this paper. As an exciting alternative, one may consider using nanoparticles consisting of magnetic Weyl semimetals, where there is a large intrinsic effective magnetic field that breaks reciprocity\cite{zhao_axion-field-enabled_2020,tsurimaki_large_2020}. Wavelength-scale objects comprised of reciprocal media can also experience a non-zero torque out of thermal equilibrium\cite{reid_photon_2017} and hence should also operate as a heat engine. Our work points to a new mechanism for harvesting thermal fluctuations.

\begin{acknowledgments}
This work is supported by the U. S. Defense Advanced Research Projects Agency (DARPA) Agreement No. HR00112090080. S. F. acknowledges useful discussions with Charles Chase. 
\end{acknowledgments}

\newpage

\setcounter{equation}{0}

\begin{center}
\textbf{Supplementary Material}
\end{center}

\section{I. Torque on a rotating gyrotropic particle}
We follow Ref.~\cite{manjavacas_vacuum_2010,manjavacas_thermal_2010} and extend the results to a nanoparticle consisting of gyrotropic materials.

For the nanoparticle shown in Fig. 1 of the main text, located at a position $r_0$, the torque along the rotation axis $\hat{z}$ is
\begin{align}
M = <p(t)\times E(r_0,t)>\cdot \hat{z}
\end{align}
where $p(t)$ is the fluctuating dipole moment, $E(r_0,t)$ is the fluctuating electric field. $<>$ denotes statistical ensemble average.
Following Ref.~\cite{manjavacas_vacuum_2010,manjavacas_thermal_2010}, we write
\begin{align}\label{torque}
M = <p^{fl}(t)\times E^{ind}(r_0,t)+p^{ind}(t)\times E^{fl}(r_0,t)>\cdot \hat{z} \equiv M_p+M_E
\end{align}
where $p^{fl}$ is the dipole in the nanoparticle, $E^{ind}$ is the field generated by $p^{fl}$, $E^{fl}$ is the field generated by the fluctuation in the vacuum region, $p^{ind}$ is the induced dipole moment in the particle as generated by $E^{fl}$. In the frequency domain, these quantities in Eq.~(\ref{torque}) satisfy:
\begin{align}\label{p_cor}
<p^{fl}(\omega)p^{fl}(\omega')>=4\pi\hbar\delta(\omega+\omega')\frac{\alpha(\omega)-\alpha^\dagger(\omega)}{2i}n_1(\omega)
\end{align}
\begin{align}\label{e_cor}
<E^{fl}(r,\omega)E^{fl}(r',\omega')>=4\pi\hbar\delta(\omega+\omega')\frac{G(r,r',\omega)-G(r',r,\omega)^\dagger}{2i}n_0(\omega)
\end{align}
\begin{align}
E^{ind}(r,\omega)=G(r,r_0,\omega)p^{fl}(\omega)
\end{align}
\begin{align}
p^{ind}(\omega)=\alpha(\omega)E^{fl}(r_0,\omega)
\end{align}
where $n_i(\omega)=\frac{1}{e^{\hbar\omega/(k_BT_i)}-1}+\frac{1}{2}=-n_i(-\omega)$ with $i\in[0,1]$, $G(r,r',\omega)$ is the free space Green's function and $\alpha$ is the frequency dependent electric polarizability of the particle which takes the form of 
\begin{align}
\alpha=\begin{pmatrix}
\alpha_{xx} & \alpha_{xy} &  0 \\
\alpha_{yx} & \alpha_{yy} & 0 \\
0 & 0 &\alpha_{zz}
\end{pmatrix}
\end{align}
As in the main text, we denote the upper $2\times 2$ principal minor of the matrix $\alpha$ by $\alpha_\perp$, that is, $\alpha_\perp=[\alpha_{xx}, \alpha_{xy}; \alpha_{yx}, \alpha_{yy}]$. We expand $\alpha_\perp$ as
\begin{align}
\alpha_\perp=\alpha_0I_2+\alpha_1\sigma_x+\alpha_2\sigma_y+\alpha_3\sigma_z,
\end{align}
where $I_2$ is the 2$\times$2 identity matrix and $\sigma_x$, $\sigma_y$, $\sigma_z$ are the standard Pauli matrices. We have 
\begin{align}
\frac{\alpha_\perp-\alpha_\perp^\dagger}{2}=\Im(\alpha_0)I_2+\Im(\alpha_1)\sigma_x+\Im(\alpha_2)\sigma_y+\Im(\alpha_3)\sigma_z
\end{align}

In Eq.~(\ref{torque})
\begin{align}
\begin{split}
M_p&=<p^{fl}(t)\times E^{ind}(r_0,t)>\cdot \hat{z}\\
&=\iint\frac{d\omega d\omega'}{4\pi^2}e^{-i(\omega+\omega')t}<p^{fl}(\omega)\times E^{ind}(r_0,\omega')>\cdot \hat{z} \\
&=\iint\frac{d\omega d\omega'}{4\pi^2}e^{-i(\omega+\omega')t}<p^{fl}(\omega)\times G(r_0,r_0,\omega')p^{fl}(\omega')> \cdot \hat{z}. \\
\end{split}
\end{align}
Here, and for the rest of the paper we denote $\int d\omega \equiv \int_ {-\infty}^\infty d\omega$. 
We note that 
\begin{align}
(A\times B)\cdot \hat{z} = \Tr(A(SB)^T)
\end{align}
where $A$ and $B$ are column vectors, and we define
\begin{align}\label{smat}
S=\begin{pmatrix}
0 & 1 & 0 \\
-1 & 0 & 0 \\
0 & 0 & 0
\end{pmatrix}.
\end{align}
$S$ satisfies $S^T=-S$.
We then have 
\begin{align}
\begin{split}
M_p
&=-\iint\frac{d\omega d\omega'}{4\pi^2}e^{-i(\omega+\omega')t}<G(r_0,r_0,\omega)p^{fl}(\omega)\times p^{fl}(\omega') >\cdot \hat{z} \\
&=-\Tr\iint\frac{d\omega d\omega'}{4\pi^2}e^{-i(\omega+\omega')t}<G(r_0,r_0,\omega)p^{fl}(\omega)(Sp^{fl}(\omega'))^T > \\
&=-\Tr\iint\frac{d\omega d\omega'}{4\pi^2}e^{-i(\omega+\omega')t}<G(r_0,r_0,\omega)p^{fl}(\omega)p^{fl}(\omega')^TS^T > \\
&=\Tr\iint\frac{d\omega d\omega'}{4\pi^2}e^{-i(\omega+\omega')t}<G(r_0,r_0,\omega)p^{fl}(\omega)p^{fl}(\omega')^TS > \\
\end{split}
\end{align}

$G(r_0,r_0,\omega)$, $S$ and are both $3\times 3$ matrices. We denote their upper $2\times 2$ principal minors with a subscript $\perp$. For instance, $G_\perp=[G_{xx}, G_{xy}; G_{yx}, G_{yy}]$. We also denote 
\begin{align}
p_\perp(\omega)=\begin{pmatrix}
p_x(\omega) \\
p_y(\omega)
\end{pmatrix}
\end{align}
for both induced and fluctuating dipole moments. 
Since $S$ is a block matrix with $S_{zz}=0$, we have 
\begin{align}\label{M_p}
\begin{split}
M_p
&=\Tr\iint\frac{d\omega d\omega'}{4\pi^2}e^{-i(\omega+\omega')t}<G_\perp(r_0,r_0,\omega)p_\perp^{fl}(\omega)p_\perp^{fl}(\omega')^TS_\perp > \\
\end{split}
\end{align}
For a static particle, using Eq.~(\ref{p_cor}), 
\begin{align}\label{M_p_static}
M_p
=\frac{\hbar}{\pi}\Tr\int{d\omega}n_1(\omega)G_\perp(r_0,r_0,\omega)\frac{\alpha_\perp(\omega)-\alpha_\perp^\dagger(\omega)}{2i}S_{\perp}
\end{align}

Similarly, the $M_E$ term in Eq.~({\ref{torque}}) can be evaluated as:
\begin{align}\label{M_E}
\begin{split}
M_E&=<p^{ind}(t)\times E^{fl}(r_0,t)>\cdot \hat{z}\\
&=\iint\frac{d\omega d\omega'}{4\pi^2}e^{-i(\omega+\omega')t}<p^{ind}(\omega)\times E^{fl}(r_0,\omega')>\cdot \hat{z} \\
&=\iint\frac{d\omega d\omega'}{4\pi^2}e^{-i(\omega+\omega')t}<\alpha(\omega)
E^{fl}(r_0,\omega)\times E^{fl}(r_0,\omega')>\cdot \hat{z} \\
&=\Tr\iint\frac{d\omega d\omega'}{4\pi^2}e^{-i(\omega+\omega')t}<\alpha(\omega)E^{fl}(r_0,\omega)E^{fl}(r_0,\omega')^T>S^T \\
&=-\Tr\iint\frac{d\omega d\omega'}{4\pi^2}e^{-i(\omega+\omega')t}<\alpha(\omega)E^{fl}(r_0,\omega)E^{fl}(r_0,\omega')^TS> \\
&=-\frac{\hbar}{\pi}\Tr\int{d\omega}n_0(\omega)\alpha(\omega)\frac{G(r_0,r_0,\omega)-G^\dagger(r_0,r_0,\omega)}{2i}S \\
&=-\frac{\hbar}{\pi}\Tr\int{d\omega}n_0(\omega)\alpha_\perp(\omega)\frac{G_\perp(r_0,r_0,\omega)-G_\perp^\dagger(r_0,r_0,\omega)}{2i}S_\perp
\end{split}
\end{align}

We now consider the torque on a rotating gyrotropic particle. We assume that the particle is rotating around the $z$-axis with an angular velocity $\Omega$. Denote $\omega_+ = \omega + \Omega$ and $\omega_-=\omega-\Omega$. The dipole moment of the particle in the lab frame can be related to that in the co-rotating frame via the Lorentz transformation as:
\begin{align}
p_z(\omega) = p_z'(\omega)
\end{align}
and

\begin{align}
\begin{pmatrix}
p_x(\omega) \\
p_y(\omega)
\end{pmatrix}
=\frac{1}{2}
\begin{pmatrix}
1 & i \\
-i & 1
\end{pmatrix}
\begin{pmatrix}
p_x'(\omega_+) \\
p_y'(\omega_+)
\end{pmatrix}
+\frac{1}{2}
\begin{pmatrix}
1 & -i \\
i & 1
\end{pmatrix}
\begin{pmatrix}
p_x'(\omega_-) \\
p_y'(\omega_-)
\end{pmatrix}
\end{align}
Similarly, the electric fields in the lab frame (the un-primed quantities) and the co-rotating frame (the primed quantities) are related as:
\begin{align}
E_z(\omega) = E_z'(\omega)
\end{align}

\begin{align}\label{transform_xy}
\begin{pmatrix}
E_x'(\omega) \\
E_y'(\omega)
\end{pmatrix}
=\frac{1}{2}
\begin{pmatrix}
1 & -i \\
i & 1
\end{pmatrix}
\begin{pmatrix}
E_x(\omega_+) \\
E_y(\omega_+)
\end{pmatrix}
+\frac{1}{2}
\begin{pmatrix}
1 & i \\
-i & 1
\end{pmatrix}
\begin{pmatrix}
E_x(\omega_-) \\
E_y(\omega_-)
\end{pmatrix}
\end{align}
Hence, in the lab frame,
\begin{align}
\begin{split}
\begin{pmatrix}
p_x(\omega) \\
p_y(\omega)
\end{pmatrix}
&=\frac{1}{2}
\begin{pmatrix}
1 & i \\
-i & 1
\end{pmatrix}
\alpha_\perp(\omega_+)\begin{pmatrix}
E_x'(\omega_+) \\
E_y'(\omega_+)
\end{pmatrix}
+\frac{1}{2}
\begin{pmatrix}
1 & -i \\
i & 1
\end{pmatrix}
\alpha_\perp(\omega_-)\begin{pmatrix}
E_x'(\omega_-) \\
E_y'(\omega_-)
\end{pmatrix} \\
&=\frac{1}{2}
\begin{pmatrix}
1 & i \\
-i & 1
\end{pmatrix}
\alpha_\perp(\omega_+)\begin{pmatrix}
E_x(\omega) \\
E_y(\omega)
\end{pmatrix}
+\frac{1}{2}
\begin{pmatrix}
1 & -i \\
i & 1
\end{pmatrix}
\alpha_\perp(\omega_-)\begin{pmatrix}
E_x(\omega) \\
E_y(\omega)
\end{pmatrix} \\
\end{split}
\end{align}
where we have used Eq.~(\ref{transform_xy}) and 
\begin{align}
\begin{pmatrix}
1 & \pm i \\
\mp i & 1
\end{pmatrix}
\begin{pmatrix}
1 & \pm i \\
\mp i & 1
\end{pmatrix}
=2\begin{pmatrix}
1 & \pm i \\
\mp i & 1
\end{pmatrix},\:
\begin{pmatrix}
1 & \pm i \\
\mp i & 1
\end{pmatrix}
\begin{pmatrix}
1 & \mp i \\
\pm i & 1
\end{pmatrix}
=0
\end{align}
We can then define an effective polarizability $\alpha_\perp^{eff}$ as
\begin{align}\label{alpha_eff}
\alpha_\perp^{eff}(\omega)=\frac{1}{2}
\begin{pmatrix}
1 & i \\
-i & 1
\end{pmatrix}
\alpha_\perp(\omega_+)
+\frac{1}{2}
\begin{pmatrix}
1 & -i \\
i & 1
\end{pmatrix}
\alpha_\perp(\omega_-)
\end{align}

We first compute the $M_p$ term in Eq.~(\ref{torque}) for the rotating particle. In Eq.~(\ref{M_p}), 
\begin{align}\label{p_rot_cor}
\begin{split}
&<p_\perp^{fl}(\omega)p_\perp^{fl}(\omega')^T>
=<[\frac{1}{2}
\begin{pmatrix}
1 & i \\
-i & 1
\end{pmatrix}
\begin{pmatrix}
p_x^{\prime fl}(\omega_+) \\
p_y^{\prime fl}(\omega_+)
\end{pmatrix}
+\frac{1}{2}
\begin{pmatrix}
1 & -i \\
i & 1
\end{pmatrix}
\begin{pmatrix}
p_x^{\prime fl}(\omega_-) \\
p_y^{\prime fl}(\omega_-)
\end{pmatrix}] \\
&\qquad \qquad [\frac{1}{2}
\begin{pmatrix}
p_x^{\prime fl}(\omega'_+) &
p_y^{\prime fl}(\omega'_+)
\end{pmatrix}
\begin{pmatrix}
1 & -i \\
i & 1
\end{pmatrix}
+\frac{1}{2}
\begin{pmatrix}
p_x^{\prime fl}(\omega'_-) &
p_y^{\prime fl}(\omega'_-)
\end{pmatrix}
\begin{pmatrix}
1 & i \\
-i & 1
\end{pmatrix}
]> \\
&=2\pi\hbar[\delta(\omega_++\omega'_+)n_1(\omega_+)(\Im(\alpha_1(\omega_+))(I-\sigma_y)\sigma_x+\Im(\alpha_3(\omega_+))(I-\sigma_y)\sigma_z) \\
&\qquad +\delta(\omega_-+\omega'_-)n_1(\omega_-)(\Im(\alpha_1(\omega_-))(I+\sigma_y)\sigma_x+\Im(\alpha_3(\omega_-))(I+\sigma_y)\sigma_z) \\
&\qquad +\delta(\omega_++\omega'_-)n_1(\omega_+)(\Im(\alpha_0(\omega_+))-\Im(\alpha_2(\omega_+)))(I-\sigma_y) \\
&\qquad +\delta(\omega_-+\omega'_+)n_1(\omega_-)(\Im(\alpha_0(\omega_-))+\Im(\alpha_2(\omega_-)))(I+\sigma_y)] \\
\end{split}
\end{align}

Plugging Eq.~(\ref{p_rot_cor}) into Eq.~(\ref{M_p}), and noting that $S_\perp = i\sigma_y$, we have 
\begin{align}
\begin{split}
M_p
&=\frac{i\hbar}{2\pi}\Tr\iint{d\omega d\omega'}e^{-i(\omega+\omega')t}G_\perp(r_0,r_0,\omega) \nonumber \\
&\quad [\delta(\omega_++\omega'_-)n_1(\omega_+)(\Im(\alpha_0(\omega_+))-\Im(\alpha_2(\omega_+)))(\sigma_y-I) \\
&\quad +\delta(\omega_-+\omega'_+)n_1(\omega_-)(\Im(\alpha_0(\omega_-))+\Im(\alpha_2(\omega_-)))(I+\sigma_y) ] \\
&=\frac{i\hbar}{2\pi}\Tr\int{d\omega}G_\perp(r_0,r_0,\omega) \\
&\quad [n_1(\omega_+)(\Im(\alpha_2(\omega_+))-\Im(\alpha_0(\omega_+))) +n_1(\omega_-)(\Im(\alpha_0(\omega_-))+\Im(\alpha_2(\omega_-))) ]
\end{split}
\end{align}
where in the first equality, the $\delta(\omega_+ +\omega'_+)$ and $\delta(\omega_- +\omega'_-)$ terms in Eq.~(\ref{p_rot_cor}) are oscillating in time and are averaged to be zero, in the second equality the $\sigma_y$ term does not contribute to the trace since $G_\perp(r_0,r_0,\omega)$ is symmetric and $\sigma_y$ is anti-symmetric.

Further, from the parity property
\begin{align}\label{parity}
G(r_0,r_0,\omega)=G^*(r_0,r_0,-\omega), \Im\alpha_0(\omega)=-\Im\alpha_0(-\omega), \Im\alpha_2(\omega)=\Im\alpha_2(-\omega),
\end{align}
and $n_i(\omega)=-n_i(-\omega)$ ($i\in[0,1]$), we find that
\begin{align}
\begin{split}
M_p
&=\frac{\hbar}{\pi}\Tr\int_0^\infty{d\omega}\Im(G_\perp(r_0,r_0,\omega)) \\
&\quad [n_1(\omega_+)(\Im(\alpha_0(\omega_+))-\Im(\alpha_2(\omega_+))) -n_1(\omega_-)(\Im(\alpha_0(\omega_-))+\Im(\alpha_2(\omega_-))) ] \\
&=\frac{4\hbar}{3\pi c^3}\int_0^\infty{d\omega}\omega^3 \\
&\quad [n_1(\omega_+)(\Im(\alpha_0(\omega_+))-\Im(\alpha_2(\omega_+))) -n_1(\omega_-)(\Im(\alpha_0(\omega_-))+\Im(\alpha_2(\omega_-))) ] \\
\end{split}
\end{align}
where we have used that 
\begin{align}\label{G0}
\Im(G_\perp(r_0,r_0,\omega))=\frac{2\omega^3}{3c^3}I_2
\end{align}
with $I_2$ being the $2\times 2$ identity matrix. 

We next compute the $M_E$ term  in Eq.~(\ref{torque}) for a rotating particle. 
From Eqs.~(\ref{M_E}) and (\ref{alpha_eff}), noting that $S_\perp=i\sigma_y$, we have
\begin{align}
\begin{split}
M_E
&=-\frac{i\hbar}{2\pi}\Tr\int{d\omega}n_0(\omega)[(\sigma_y-I)\alpha(\omega_+)+(\sigma_y+I)\alpha(\omega_-)]\Im(G_\perp(r_0,r_0,\omega)) \\
&=-\frac{i\hbar}{2\pi}\Tr\int{d\omega}n_0(\omega)(\alpha_2(\omega_+)-\alpha_0(\omega_+)+\alpha_2(\omega_-)+\alpha_0(\omega_-))\Im(G_\perp(r_0,r_0,\omega)) \\
\end{split}
\end{align}
With Eqs.~(\ref{parity}) and (\ref{G0}), 
\begin{align}
\begin{split}
M_E
&=\frac{4\hbar}{3\pi c^3}\int_0^\infty{d\omega}\omega^3 n_0(\omega)(\Im(\alpha_2(\omega_+))-\Im(\alpha_0(\omega_+))+\Im(\alpha_2(\omega_-))+\Im(\alpha_0(\omega_-))) \\
\end{split}
\end{align}

We now compute the total torque $M=M_p+M_E$. 
\begin{align}
\begin{split}
M=&-\frac{4\hbar}{3\pi c^3}\int_0^\infty{d\omega}\omega^3 \Im(\alpha_0(\omega_-))(n_1(\omega_-)-n_0(\omega)) \\
&+\frac{4\hbar}{3\pi c^3}\int_0^\infty{d\omega}\omega^3 \Im(\alpha_0(\omega_+))(n_1(\omega_+)-n_0(\omega)) \\
&-\frac{4\hbar}{3\pi c^3}\int_0^\infty{d\omega}\omega^3 \Im(\alpha_2(\omega_-))(n_1(\omega_-)-n_0(\omega)) \\
&-\frac{4\hbar}{3\pi c^3}\int_0^\infty{d\omega}\omega^3 \Im(\alpha_2(\omega_+))(n_1(\omega_+)-n_0(\omega)) \\
\end{split}
\end{align}
By playing with the integration limits,
\begin{align}
\begin{split}
M=-\frac{4\hbar}{3\pi c^3}\int_{-\infty}^\infty{d\omega}\omega^3 (\Im (\alpha_0(\omega_-))+\Im (\alpha_2(\omega_-)))(n_1(\omega_-)-n_0(\omega)) \\
\end{split}
\end{align}

\section{II. Radiation power}
The power radiated by the particle is
\begin{align}
P = -<E^{ind}(r_0,t)\cdot\frac{\partial p^{fl}(t)}{\partial t}>-<E^{fl}(r_0,t)\cdot \frac{\partial p^{ind}(t)}{\partial t}> \equiv P_p+P_E
\end{align}

Here we first compute the radiated power $P_\perp$ from the dipole components in the $x$-$y$ plane. 

\begin{align}\label{P_p}
\begin{split}
P_{\perp p} &= -<E_\perp^{ind}(t)\cdot\frac{\partial p_\perp^{fl}(t)}{\partial t}> \\
&= \Tr\iint_{-\infty}^\infty\frac{d\omega d\omega'}{4\pi^2}e^{-i(\omega+\omega')t}(i\omega')G_\perp(r_0,r_0,\omega)<p_\perp^{fl}(\omega)(p_\perp^{fl}(\omega'))^T> \\
\end{split}
\end{align}
Plugging Eq.~(\ref{p_rot_cor}) into Eq.~(\ref{P_p}), we have 
\begin{align}
\begin{split}
P_{\perp p}
&=-\frac{i\hbar}{2\pi}\Tr\int_{-\infty}^\infty{d\omega}\omega G_\perp(r_0,r_0,\omega)[n_1(\omega_+)(\Im(\alpha_0(\omega_+))-\Im(\alpha_2(\omega_+)))(I-\sigma_y) \\
&\qquad \qquad \qquad \qquad +n_1(\omega_-)(\Im(\alpha_0(\omega_-))+\Im(\alpha_2(\omega_-)))(I+\sigma_y)] \\
&=-\frac{i\hbar}{2\pi}\Tr\int_{-\infty}^\infty{d\omega}\omega G(r_0,r_0,\omega)[n_1(\omega_+)(\Im(\alpha_0(\omega_+))-\Im(\alpha_2(\omega_+))) \\
&\qquad \qquad \qquad \qquad +n_1(\omega_-)(\Im(\alpha_0(\omega_-))+\Im(\alpha_2(\omega_-)))] \\
\end{split}
\end{align}
With the parity property Eq.~(\ref{parity}), 
\begin{align}
\begin{split}
P_{\perp p}
&=\frac{\hbar}{\pi}\Tr\int_{0}^\infty{d\omega}\omega \Im G(r_0,r_0,\omega)[n_1(\omega_+)(\Im(\alpha_0(\omega_+))-\Im(\alpha_2(\omega_+))) \\
&\qquad +n_1(\omega_-)(\Im(\alpha_0(\omega_-))+\Im(\alpha_2(\omega_-)))] \\
&=\frac{4\hbar}{3\pi c^3}\int_{0}^\infty{d\omega}\omega^4[n_1(\omega_+)(\Im(\alpha_0(\omega_+))-\Im(\alpha_2(\omega_+))) \\
&\qquad +n_1(\omega_-)(\Im(\alpha_0(\omega_-))+\Im(\alpha_2(\omega_-)))] \\
\end{split}
\end{align}

We now compute $p_{\perp E}$, we have 
\begin{align}\label{P_E}
\begin{split}
p_{\perp E}&=-<E_\perp^{fl}(r_0,t)\cdot \frac{\partial p_\perp^{ind}(t)}{\partial t}> \\
&= \Tr\iint_{-\infty}^\infty\frac{d\omega d\omega'}{4\pi^2}e^{-i(\omega+\omega')t}(i\omega')E_\perp^{fl}(r_0,\omega)(p_\perp^{ind}(\omega'))^T \\
&= \Tr\iint_{-\infty}^\infty\frac{d\omega d\omega'}{4\pi^2}e^{-i(\omega+\omega')t}(i\omega')E_\perp^{fl}(r_0,\omega)(\alpha^{eff}(\omega')E_\perp^{fl}(r_0,\omega'))^T \\
&= \Tr\iint_{-\infty}^\infty\frac{d\omega d\omega'}{4\pi^2}e^{-i(\omega+\omega')t}(i\omega')E^{fl}(r_0, \omega)(E^{fl}(r_0, \omega'))^T(\alpha_\perp^{eff}(\omega'))^T \\
\end{split}
\end{align}
Plugging Eq.~(\ref{alpha_eff}) to Eq.~(\ref{P_E}), using Eq.~(\ref{e_cor}), and noting that $\alpha_\perp^{eff}(-\omega)=(\alpha_\perp^{eff}(\omega))^*$, we have 
\begin{align}
\begin{split}
p_E^{rad}
&= -\frac{2i\hbar}{3\pi c^3}\Tr\int_{-\infty}^\infty{d\omega}\omega^4 n_0(\omega)(\alpha^{eff}(\omega))^\dagger \\
&= -\frac{i\hbar}{3\pi c^3}\Tr\int_{-\infty}^\infty{d\omega}\omega^4 n_0(\omega)[(1-\sigma_y)\alpha^\dagger(\omega_+)+(1+\sigma_y)\alpha^\dagger(\omega_-)] \\
&= -\frac{2i\hbar}{3\pi c^3}\Tr\int_{-\infty}^\infty{d\omega}\omega^4 n_0(\omega)(\alpha_0^*(\omega_+)-\alpha_2^*(\omega_+)+\alpha_0^*(\omega_-)+\alpha_2^*(\omega_-)) \\
&= -\frac{4\hbar}{3\pi c^3}\int_{0}^\infty{d\omega}\omega^4 n_0(\omega)(\Im\alpha_0(\omega_+)-\Im\alpha_2(\omega_+)+\Im\alpha_0(\omega_-)+\Im\alpha_2(\omega_-)) \\
\end{split}
\end{align}

We now compute the total $P_\perp$. 
\begin{align}
\begin{split}
P_\perp&=P_{\perp p}+P_{\perp E}\\
&=\frac{4\hbar}{3\pi c^3}\int_{0}^\infty{d\omega}\omega^4\Im\alpha_0(\omega_-)(n_1(\omega_-)-n_0(\omega))\\
&+\frac{4\hbar}{3\pi c^3}\int_{0}^\infty{d\omega}\omega^4\Im\alpha_0(\omega_+)(n_1(\omega_+)-n_0(\omega))\\
&+\frac{4\hbar}{3\pi c^3}\int_{0}^\infty{d\omega}\omega^4\Im\alpha_2(\omega_-)(n_1(\omega_-)-n_0(\omega))\\
&-\frac{4\hbar}{3\pi c^3}\int_{0}^\infty{d\omega}\omega^4\Im\alpha_2(\omega_+)(n_1(\omega_+)-n_0(\omega))
\end{split}
\end{align}
By playing with the integral limits, we have 
\begin{align}
\begin{split}
P_\perp
&=\frac{4\hbar}{3\pi c^3}\int_{-\infty}^\infty{d\omega}\omega^4(\Im(\alpha_0(\omega_-))+\Im(\alpha_2(\omega_-)))(n_1(\omega_-)-n_0(\omega))
\end{split}
\end{align}
Likewise, the radiated power from the dipole along the $z$-direction can be found to be \cite{manjavacas_vacuum_2010,manjavacas_thermal_2010}
\begin{align}
\begin{split}
P_\parallel=
\frac{2\hbar}{3\pi c^3}\int_{-\infty}^\infty{d\omega}\omega^4\Im\alpha_{zz}(\omega)(n_1(\omega)-n_0(\omega))\\
\end{split}
\end{align}

\section{III. Proof of $\omega g_\perp(\omega)\geq 0$}
The polarizability $\alpha$ is a linear response function of a static and passive system. Hence $\omega\frac{\alpha(\omega)-\alpha^\dagger(\omega)}{2}$ must be a semi-positive definite matrix. Its principle minor 
\begin{align}
\omega\frac{\alpha_\perp(\omega)-\alpha_\perp^\dagger(\omega)}{2}=\omega(\Im(\alpha_0)I+\Im(\alpha_1)\sigma_x+\Im(\alpha_2)\sigma_y+\Im(\alpha_3)\sigma_z)
\end{align}
The determinant of this matrix must be non-negative, 
\begin{align}\label{det}
\omega^2(\Im^2(\alpha_0)-\Im^2(\alpha_1)-\Im^2(\alpha_2)-\Im^2(\alpha_3)) \geq 0
\end{align}
Also, the first diagonal element of the matrix must be non-negative, 
\begin{align}\label{diagonal}
\omega(\Im(\alpha_0)+\Im(\alpha_3)) \geq 0. 
\end{align}
Combining Eqs.~(\ref{det}) and (\ref{diagonal}), we have
\begin{align}
\omega(\Im(\alpha_0)+\Im(\alpha_i)) \geq 0, \, i=1,2,3. 
\end{align}
In particular, 
\begin{align}
\omega g_\perp(\omega) =
\omega(\Im(\alpha_0)+\Im(\alpha_2)) \geq 0.
\end{align}

\section{IV. Low loss limit}
For a nanoparticle with a dielectric permittivity as described by Eq.~(17) of the main text with $\epsilon_\infty=1$, $\epsilon_{zz}=2$, its polarizability $\alpha$ has the form:
\begin{align}
\begin{split}
\Im(\alpha_0+\alpha_2)
&= \frac{9\gamma\omega\omega_p^2}{(\omega_p^2-3\omega^2+3\omega\omega_c)^2+(3\omega\gamma)^2}
\end{split}
\end{align}

$\omega_p^2-3\omega^2+3\omega\omega_c=0$ has two roots, 
\begin{align}
\begin{split}
\omega_{01,02}=\frac{3\omega_c\pm \sqrt{9\omega_c^2+12\omega_p^2}}{6}
\end{split}
\end{align}

In the limit $\gamma\to 0$,
\begin{align}
\begin{split}
\Im(\alpha_0+\alpha_2)&=\pi\delta(\omega-\omega_{01})\frac{\omega_p^2}{2\omega_{01}-\omega_c}+\pi\delta(\omega-\omega_{02})\frac{\omega_p^2}{2\omega_{02}-\omega_c}\\
&=\frac{3\pi\omega_p^2}{\sqrt{9\omega_c^2+12\omega_p^2}}(\delta(\omega-\omega_{01})-\delta(\omega-\omega_{02}))\\
\end{split}
\end{align}
In the main text, we use $\omega_1=\omega_{01}$ and $\omega_2=-\omega_{02}$.

\end{document}